# Gate tunability of highly efficient spin-to-charge conversion by spin Hall effect in graphene proximitized with WSe$_2$

Franz Herling[1], C. K. Safeer[1], Josep Ingla-Aynés[1], Nerea Ontoso[1], Luis E. Hueso[1,2], Fèlix Casanova[1,2,*]

[1]CIC nanoGUNE BRTA, 20018 Donostia-San Sebastian, Basque Country, Spain.
[2]IKERBASQUE, Basque Foundation for Science, 48013 Bilbao, Basque Country, Spain.

*E-mail: f.casanova@nanogune.eu

## Abstract

The proximity effect opens ways to transfer properties from one material into another and is especially important in two-dimensional materials. In van der Waals heterostructures, transition metal dichalcogenides (TMD) can be used to enhance the spin-orbit coupling of graphene leading to the prediction of gate controllable spin-to-charge conversion (SCC). Here, we report for the first time and quantify the SHE in graphene proximitized with WSe$_2$ up to room temperature. Unlike in other graphene/TMD devices, the sole SCC mechanism is the spin Hall effect and no Rashba-Edelstein effect is observed. Importantly, we are able to control the SCC by applying a gate voltage. The SCC shows a high efficiency, measured with an unprecedented SCC length larger than 20 nm. These results show the capability of two-dimensional materials to advance towards the implementation of novel spin-based devices and future applications.

The integration of spintronic devices into existing electronic technology will strongly depend on the all-electrical control of spin currents, with a crucial role being played by the interconversion between charge currents and spin currents. The latter can be achieved by the (inverse) spin Hall effect [(I)SHE] in bulk conductors[1–3], as well as by the (inverse) Rashba-Edelstein effect [(I)REE] in two-dimensional (2D) systems and interfaces[4], allowing ferromagnet(FM)-free electrical generation and detection of spin currents. While the experimental observation of the (I)SHE and (I)REE have been successful in different systems[3,5,6], the transition from the laboratory to industrial applications will require careful device design and material choice to achieve large enough signals for practical implementation[7–10].

Since the first mechanical exfoliation of graphene[11], the library of two-dimensional (2D) materials has grown[12,13], with a plethora of materials that possess a wide range of properties that are complementary to those of graphene. Deterministic transfer methods[14] allow to combine these properties by stacking different 2D materials into a van der Waals heterostructure[15–17]. New properties such as magnetism, superconductivity or spin-orbit coupling can also be induced in graphene by proximity[18–21], leading to new functionalities[22–24]. In particular, the combination of graphene and semiconducting transition metal dichalcogenides (TMDs) with strong spin-orbit coupling (SOC) is a promising platform to study a variety of spin-dependent phenomena. For instance, using the tunable conductivity of a graphene/TMD heterostructure, an electrical spin field-effect switch has already been realized[25,26]. More interestingly, a large SOC can be imprinted by proximity from the TMD into graphene[27,28] leading to the presence of weak antilocalization[29–

[34], spin lifetime anisotropy[35–37], (I)SHE[31,38] and (I)REE[39–41]. While the previous measurements claiming SHE in graphene used a non-local Hall bar geometry[42–44], where a variety of non-spin-related effects can contribute and make an interpretation difficult[45–49], the SHE was first unambiguously reported in graphene/$MoS_2$[38] and the REE later in graphene/$WS_2$[41]. Theoretical calculations show that the proximity SOC can be tuned by a gate voltage[23,50] which in the case of $WSe_2$ could lead to larger spin Hall angles in the electron-doped regime of graphene[31] and in general would lead towards an electrically controllable spin-to-charge conversion (SCC) device.

Here, we report for the first time the observation of the SHE in graphene proximitized with $WSe_2$. In contrast to other graphene/TMD heterostructures[38,41], the IREE does not contribute to the SCC. Importantly, the SCC signal can be amplified and turned off by an applied back-gate voltage. The amplified SCC signal is up to eleven times larger than our previously reported results in devices with proximitized graphene/$MoS_2$[38] due to a highly efficient conversion with a SCC length of up to 41 nm (with a lower limit of 20 nm), six (three) times larger than the largest value reported[51]. The high SCC efficiency combined with the extra functionality of controlling the SCC with a gate voltage thus makes this van der Waals heterostructure a promising system for the creation and detection of pure spin currents in applications such as spin-orbit logic[8,10] or electrical manipulation of magnetic memories[52–54].

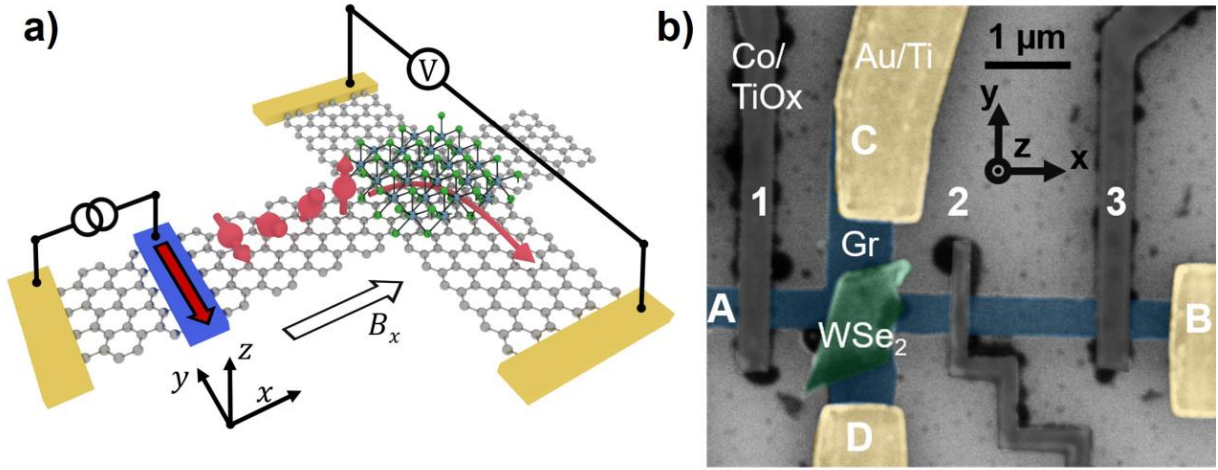

**Figure 1. a)** Sketch of the proximity-induced ISHE in graphene and the device used to measure it. The precession of the spins and the SCC is sketched. **b)** False-colored scanning electron microscopy image of the device showing the labeling of the Co/$TiO_x$ (numbers) and Au/Ti (letters, A outside of the image) contacts used for the transport measurements. The metallic Au/Ti and the FM Co/$TiO_x$ contacts enable us to measure spin transport in a reference pristine graphene channel (LSV between electrodes 2 and 3) and spin transport and ISHE in a $WSe_2$-proximitized graphene Hall bar (LSV between electrodes 1 and 2).

Our device was carefully designed to measure the (I)SHE in TMD-proximitized graphene (see sketch in Figure 1a) as well as the (I)REE[38]. The device was prepared by placing a multilayer $WSe_2$ flake by dry transfer on a trilayer graphene flake and patterning it into a Hall bar structure. Metallic electrical contacts (Au/Ti) and FM electrodes (Co) with a resistive barrier ($TiO_x$) allow for electrical spin injection and detection. The final device is shown in Figure 1b. Low-noise electrical measurements were performed while applying an in-plane magnetic field either along the $x$- or $y$- axis at temperatures between 10 and 300 K. The transport in our device is diffusive. See Note S2 for details on the device fabrication and measurements.

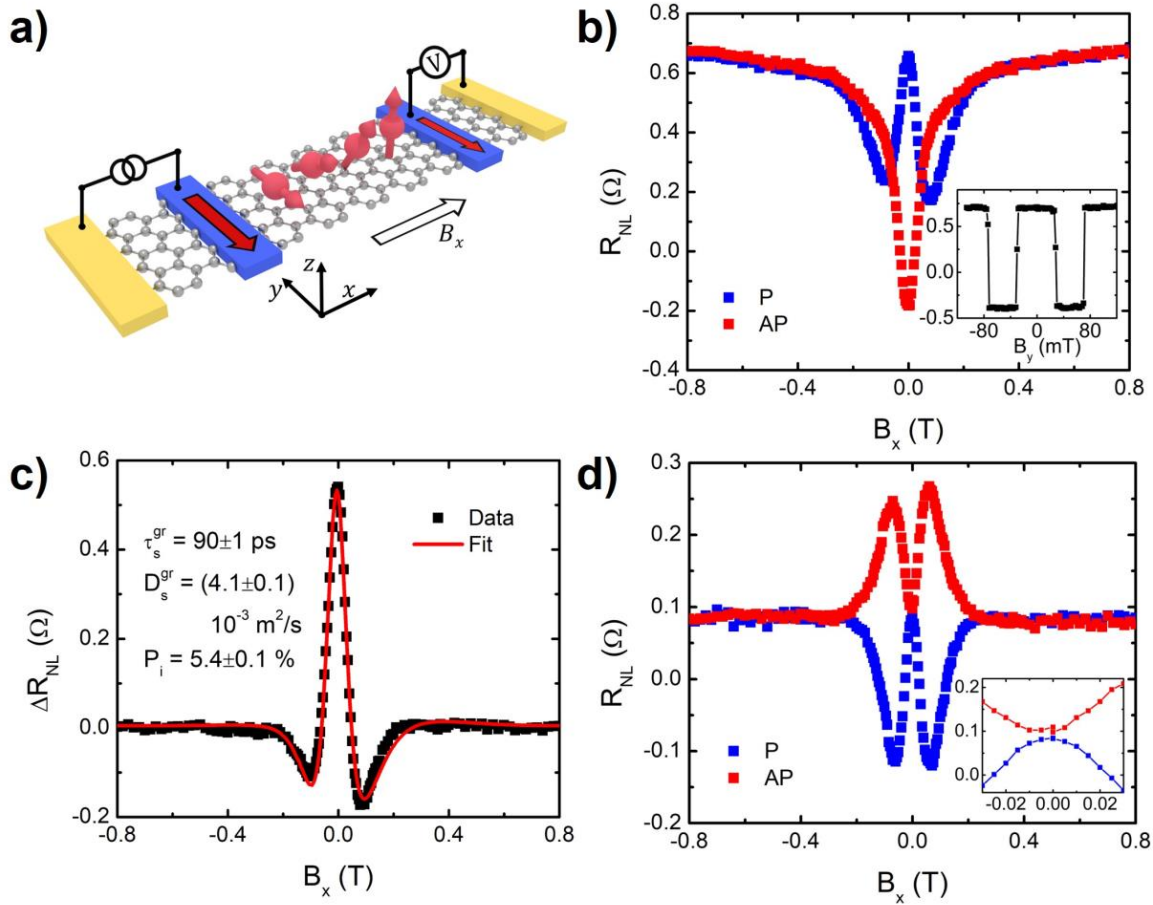

**Figure 2.** Spin transport characterization at 100 K for pristine graphene and WSe$_2$-proximitized graphene. **a)** Measurement configuration for the Hanle precession measurement showing charge current and voltage terminals and magnetic field direction. The precession of the spin polarization is sketched. **b)** Non-local resistance measured across the reference LSV (voltage 2-A and current 3-B) as a function of an in-plane magnetic field parallel to the graphene channel ($B_x$) while the injecting and detecting Co electrodes are in the parallel (blue) and antiparallel (red) magnetization configurations. Inset: The same measurement with the magnetic field parallel to the FM electrode ($B_y$). A positive spin signal of ~0.55 Ω is obtained. **c)** $\Delta R_{NL} = (R_{NL}^P - R_{NL}^{AP})/2$, obtained from the two curves in b, as a function of the magnetic field $B_x$. The red solid line is a fit of the data to the 1D diffusion equation. The extracted parameters are shown as well. **d)** Non-local resistance across the graphene/WSe$_2$ region (voltage 2-B and current 1-A) as a function of an in-plane magnetic field parallel to the graphene channel ($B_x$) while the injecting and detecting Co electrodes are in the parallel (blue) and antiparallel (red) magnetization configurations. Inset: Zoom of the measurement at low magnetic field.

The device design enables us to study one lateral spin valve (LSV) of pristine graphene and one LSV with a graphene-WSe$_2$ heterostructure in the center, using FM electrodes. Applying a charge current ($I_c$) through the Co/TiOx contacts leads to a spin accumulation in the graphene beneath the electrode, which diffuses in both directions through the 2D channel and can be measured as a non-local voltage ($V_{NL}$) across the interface between the second FM electrode and graphene (see Figure 2a). To study the spin injection and the spin transport properties of the pristine graphene, we measured the non-local resistance ($R_{NL} = V_{NL}/I_c$) for the reference LSV by applying current between contact 3 and B and detecting the voltage between contact 2 and A. The measured $R_{NL}$ changes with the relative orientation of the magnetization of the different electrodes [parallel (P) and antiparallel (AP)]. This change can be measured by sweeping the magnetic field along the easy axis of the ferromagnet ($B_y$) and is defined as the spin signal[55,56]. The measurement at 100 K can be seen in the inset of Figure 2b. The magnetizations of the electrodes switch at different coercive fields due to different shape anisotropy, which makes the P and AP states clearly visible and controllable by the proper $B_y$ history.

Setting the sample in one of those two states and applying a magnetic field along the hard axis of the ferromagnet ($B_x$) parallel to the channel leads to the precession of the injected spins around this axis. Measuring the non-local resistance for parallel ($R_{NL}^P$) and antiparallel ($R_{NL}^{AP}$) configurations as a function of magnetic field leads to the so-called symmetric Hanle precession curves[55]. Fitting the difference between those two curves ($\Delta R_{NL} = (R_{NL}^P - R_{NL}^{AP})/2$) to a 1D spin diffusion equation[57] enables us to extract the spin transport properties, i.e., the spin diffusion constant ($D_S^{gr}$), the spin lifetime ($\tau_S^{gr}$) and the spin polarization of the Co/graphene interface ($P_i$). The measurement at 100 K is plotted in Figure 2b and the corresponding fit, together with the extracted parameters, in Figure 2c. The oscillation and decay of the spin signal can be explained by the precession, diffusion and relaxation of the spins in the graphene channel. As the FM electrodes have a finite width, the pulling of the magnetization into the direction of the magnetic field, which is complete for $B_x > 0.3$ T, has been considered for the fitting (see Note S8 for details).

The same measurement was performed across the proximitized graphene region in the second LSV, applying current from contact 1 to A and detecting voltage between contact 2 and B. The resulting plot as a function of $B_x$ can be seen in Figure 2d. As theoretically predicted[37] and already experimentally observed[35,36], the enhanced SOC by proximity effect leads not only to an enhanced spin relaxation compared to the pristine graphene, but also to a large anisotropy between the in- and out-of-plane spin lifetimes ($\tau_\parallel^{gr/TMD}$ and $\tau_\perp^{gr/TMD}$ respectively). This shows up in the Hanle precession curves as a suppression of the spin signal at low fields when the spins are polarized in-plane. As the magnetic field increases, the injected spins precess out of the sample plane and acquire a lifetime which is a combination of $\tau_\parallel^{gr/TMD}$ and $\tau_\perp^{gr/TMD}$. This leads to a sign change of $\Delta R_{NL}$ and the observation of enhanced shoulders when compared with the zero-field value[35–37]. This typically allows the determination of the two spin lifetimes from the experimental data by fitting it to the solution of the anisotropic Bloch equation[35]. However, our data, while clearly showing all the other signatures of anisotropic spin transport, misses the characteristic crossing of $R_{NL}^P$ and $R_{NL}^{AP}$ at low fields (see the inset in Figure 2d), preventing us from determining $\tau_\parallel^{gr/TMD}$ and $\tau_\perp^{gr/TMD}$. It also leads to a negative sign of the spin signal at zero field. The missing crossing is a surprising result that is not expected as the enhanced shoulders that we observe are already a consequence of the out-of-plane precession which should lead to the reversal of the in-plane spin precession in this field range. We discuss the possible origin in Note S3. Whereas the shoulders show that the out-of-plane spin signal is enhanced (-0.2 Ω) and much larger than the in-plane one (-10 mΩ), it is still smaller than in the pristine graphene LSV, where we obtained 0.55 Ω (half the difference between P and AP state).

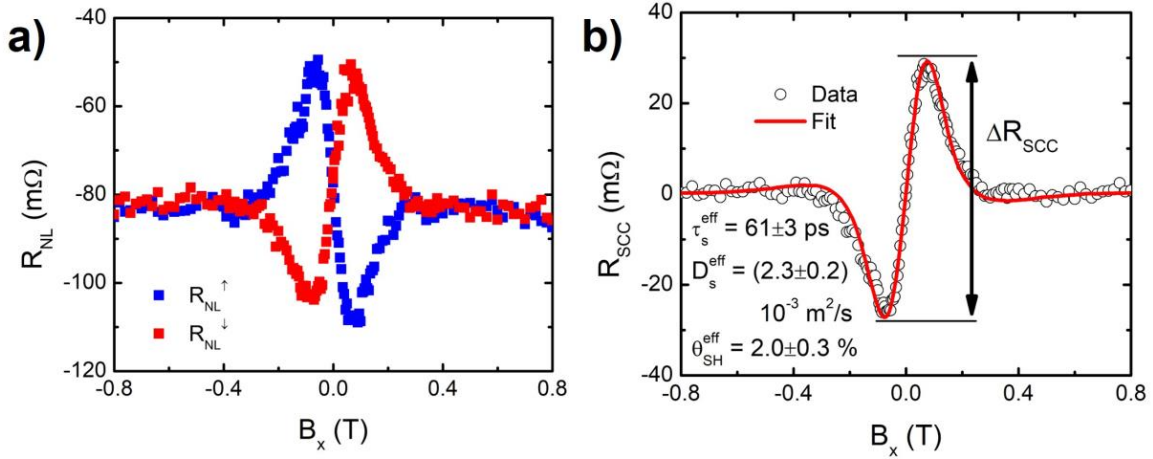

**Figure 3.** Spin-to-charge conversion measurement at 100 K. **a)** Non-local spin-to-charge conversion curves obtained by measuring across the graphene/WSe$_2$ Hall bar (current 1-A and voltage C-D). The magnetic field is applied along the in-plane hard axis direction ($B_x$) for initial positive ($R_{NL}^\uparrow$, blue) and negative ($R_{NL}^\downarrow$, red) magnetization directions of the FM electrodes. **b)** Net antisymmetric Hanle signal (open circles) obtained by subtracting the two curves ($R_{SCC} = (R_{NL}^\uparrow - R_{NL}^\downarrow)/2$) in panel a. The red solid line is a fit of the data to the diffusion equation with the extracted parameters. The definition of $\Delta R_{SCC}$ is shown with the black arrow.

The observed spin lifetime anisotropy in our symmetric Hanle curves is a fingerprint of the induced SOC in graphene by proximity with WSe$_2$[35–37]. Such a spin-orbit proximity in the graphene/WSe$_2$ region is also expected to lead to a sizable SHE, even though the intervalley scattering leading to anisotropy has been predicted to be detrimental to the SHE[40]. We used the following configuration to study the ISHE in our device: we inject the spin current into graphene by applying a charge current $I_C$ from contact 1 to A, which diffuses to both sides of the graphene channel, reaches the proximitized graphene region and is converted into a perpendicular flowing charge current that we measure as a voltage $V_{NL}$ along the graphene Hall bar with the Au/Ti contacts C and D. Due to the symmetry of the ISHE, only spins that are polarized out-of-plane, perpendicular to the direction of the spin and charge currents, will be converted. It should be noted that the device can also detect SCC due to IREE. In contrast to the SHE, the IREE will only convert spin currents that are polarized along the $x$-axis into a transverse charge. To achieve an out-of-plane spin current, an in-plane magnetic field (along the $x$-axis) is applied, that precesses the spins from the $y$-axis (parallel to the magnetic easy axis of the FM electrode) towards the $z$-axis (out-of-plane). Reversal of the magnetic field leads to a sign change in the $z$-component of the spin accumulation and, therefore in $V_{NL}$ across the graphene Hall bar and in the normalized signal $R_{NL}$. We measure the same baseline signal for in-plane polarized spin at zero fields and, when the magnetization of the FM electrode is in $x$-direction, for high fields. When the precession angle is 90° at finite, low fields, a maximum number of spins are converted, and we measure a maximum (or minimum) signal. This leads to an antisymmetric Hanle precession curve. For the two cases of initial magnetization of the Co electrode along the $y$-direction ($R_{NL}^{\uparrow}$ for positive and $R_{NL}^{\downarrow}$ for negative magnetization along the easy axis) the antisymmetric Hanle curve is reversed, as expected from the precession of spins with opposite polarization[38,40,58]. The measurement at 100 K can be seen in Figure 3a. The Onsager reciprocity, where a charge current through the graphene/WSe₂ region (along the $y$-axis) gives rise to a transverse spin current (along the $x$-axis) due to the direct SHE, is also confirmed in our device (shown in Note S4). Finally, the $R_{NL}^{\uparrow(\downarrow)}$ curve changes sign with reversing the spin current direction, which further confirms the proximity-induced ISHE in graphene as the source of the SCC (see Note S5). Results in Fig. 3a also confirm that no IREE is present in our SCC signal, as it does not switch between positive and negative high fields, when the applied magnetic field pulls and saturates the magnetization of the FM electrode along the $x$-axis and the injected spins are thus polarized in this direction[38].

Similar to the symmetric Hanle curves, the difference between the two antisymmetric Hanle precession curves, $R_{SCC} = (R_{NL}^{\uparrow} - R_{NL}^{\downarrow})/2$, gives the net signal that can be fitted to the solution of the Bloch equation[57], as shown in Figure 3b for the case of 100 K alongside the fitted parameters. We extract an effective spin lifetime ($\tau_s^{eff}$), an effective spin diffusion constant ($D_s^{eff}$) and an effective spin polarization ($P_i^{eff}$). As we now detect the spin current via the SCC in the proximitized graphene/WSe₂ region and not with a FM electrode, $P_i$ of the detector is replaced by the spin Hall angle $\theta_{SH}^{eff}$ and thus $P_i^{eff} = \sqrt{P_i \theta_{SH}^{eff}}$. Assuming the same $P_i$ for the injector as the one obtained from the electrode pair of the reference LSV, we can calculate the value of $\theta_{SH}^{eff}$. However, because the sign of $P_i$ is not known, the sign of $\theta_{SH}^{eff}$ cannot be determined.

In our model, the spin transport for the ISHE measurements is described with a single set of effective parameters ($\tau_s^{eff}$ and $D_s^{eff}$). This implies that the graphene/WSe₂ and the adjacent pristine graphene regions have the same spin transport parameters, which define the spin diffusion length ($\lambda_s^{eff} = \sqrt{\tau_s^{eff} D_s^{eff}}$). This approximation was necessary to perform the quantitative analysis, as we were unable to extract the in-plane and out-of-plane spin lifetimes of the proximitized graphene/WSe₂ region from the symmetric Hanle curves. Since $\lambda_s^{gr}$ of the pristine graphene is expected to be larger than $\lambda_{\perp}^{gr/TMD}$, the out-of-plane spin diffusion length of the graphene/WSe₂ region [35,36], our approximation likely leads to $\lambda_s^{eff} > \lambda_{\perp}^{gr/TMD}$, which in turn leads to an underestimation of $\theta_{SH}^{eff}$. Since the product of both parameters, $\theta_{SH}^{eff} \lambda_s^{eff}$ (SCC length), is in fact a better quantity to estimate the conversion efficiency[9,59], we need to consider whether the two effects can compensate each other. In Note S11, we discuss this compensation in more detail and show that we are slightly overestimating the $\theta_{SH}^{eff} \lambda_s^{eff}$ product, by up to a factor of 2 as the upper limit.

In a next step, we measured the temperature dependence of the symmetric (spin transport) and antisymmetric (spin-to-charge conversion) Hanle curves between 10 K and 300 K. The spin transport measurements at different temperatures are shown in Note S6. The ISHE measurements at different representative temperatures with the corresponding fits are plotted in Figure 4a. We note that the SCC signal, $\Delta R_{SCC}$, defined as the difference between the minimum and maximum of $R_{SCC}$, increases with decreasing temperature (inset in Figure 4a). One contributing factor for this trend is the increasing sheet resistance of the graphene channel, that increases roughly by 40% from 300 K to 10 K (see Note S2). Also, $\lambda_s^{eff}$ and $\theta_{SH}^{eff}$ are slightly increasing with decreasing temperature, which leads to more spin current reaching the proximitized area under the WSe$_2$ flake and being converted there more efficiently at lower temperatures (see Note S12 for a list of the fitted parameters at all temperatures).

As a final experimental characterization step, we measured the back-gate voltage, $V_{bg}$, dependence of the symmetric and antisymmetric Hanle curves at 100 K. For the ISHE measurement, the resulting data together with the fits can be seen in Figure 4b and the symmetric Hanle curves in Note S6.

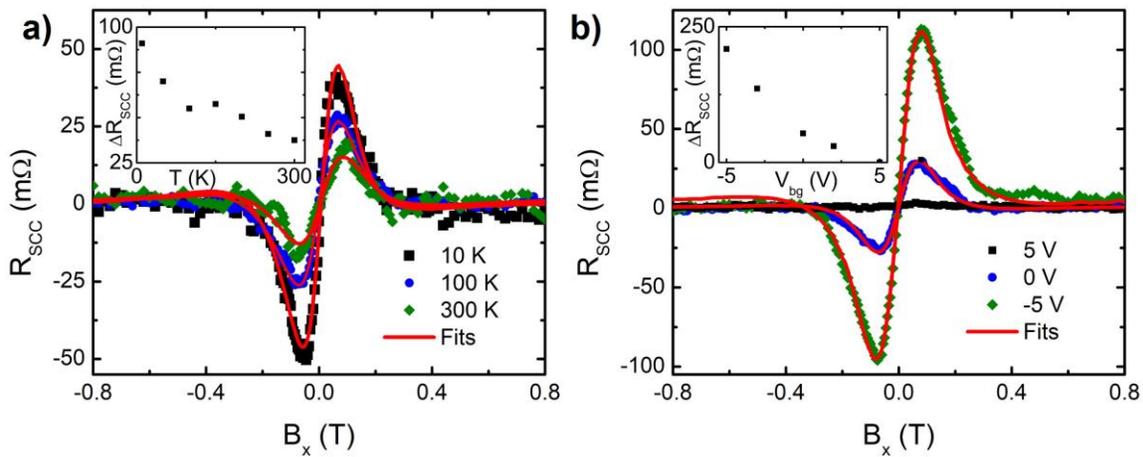

**Figure 4.** Net antisymmetric Hanle signals measured at **a)** different temperatures and zero back-gate voltage and **b)** different back-gate voltages and 100 K. The scatter plots are the experimental data, the red solid lines are fits to the data. Inset for a) $\Delta R_{SCC}$ as function of temperature at zero gate voltage and b) as function of gate voltage at 100 K.

The back-gated measurements show that the SCC signal can be increased by 400% by applying -5 V and completely suppressed for 5 V gate voltage (see inset in Fig. 4b). This gate voltage range translates into charge carrier density values from $7.2 \times 10^{11}$ cm$^{-2}$ to the charge neutrality point. The strong variation of the SCC signal cannot be explained by the change in resistance of the graphene channel, as it decreases for negative gate voltages (see Note S10), or by the effective spin diffusion length, which varies only slightly when applying positive gate voltages (see Note S13 for a list of the fitted parameters at all gate voltages). However, the estimated $\theta_{SH}^{eff}$ scales with the SCC signal and increases to 8.4% for −5 V gate voltage, whereas at 5 V it decreases below 0.2%, that we estimate as an upper limit due to the noise level. Therefore, we conclude that the gate voltage directly controls the SCC.

The gate tunability of the spin Hall effect in graphene proximitized by a TMD has been theoretically predicted, where a sign change is expected around the charge neutrality point[31]. Our gate voltage range limitation (due to a leakage current through the gate dielectric) prevented us from crossing the charge neutrality point to observe the sign change. Because of this, we cannot rule out that the suppression (amplification) of the SCC signal arises from an increased (decreased) spin absorption into the WSe$_2$ flake if the applied back-gate voltage strongly modifies the resistance of WSe$_2$ in this range. In this scenario, the largest estimated $\theta_{SH}^{eff}$ (8.4% at −5 V) would be a lower limit. In either case, though, a large tunability of the SCC signal is achieved with a back-gate voltage, an extra functionality that opens new possibilities in spin-orbit-based logic or memory.

In agreement with other experimental studies of the proximity effect of TMDs in graphene[38,41], the measured $\theta_{SH}^{eff}$ is larger than the theoretical calculation by tight-binding models[31] (from which a maximum value of 1.1% is extracted in the hole-doped regime, assuming our experimental resistance), suggesting that extrinsic sources of spin-dependent scattering such as vacancies or impurities might also be relevant in these heterostructures. It should be noted that the theoretical calculation is done for ideal monolayer graphene/monolayer TMD systems and discrepancies could therefore occur in thicker samples. However, as the proximity effect will strongly decay over distance, the SCC will mainly occur in the graphene layer adjacent to the TMD and the theoretical model should be a good approximation. As we have no control of the crystallographic alignment of the graphene and TMD flake, the twist angle between the two could also lead to a deviation from the theoretical model, which assumes a quasi-commensurate structure.

In contrast to Ref. 41, the SCC signal is solely due to ISHE, as we do not observe IREE at any temperature or gate voltage that would be visible as an "S-shaped" background in the antisymmetric Hanle measurements[38,41]. From the noise level of our background, we estimate the REE efficiency $\alpha_{RE}$ to be < 0.05%. Our results suggest that the valley-Zeeman SOC induced in graphene, main responsible of the SHE, dominates over the Rashba SOC, which generates the REE. Experimentally, the same has been found in weak antilocalization measurements of graphene/WSe$_2$ and WS$_2$[32,33,60]. The valley-Zeeman term originates in the broken sublattice symmetry of the TMD, which is imprinted into the graphene by proximity and spin polarizes the bands out of the plane with opposite orientation in the K and K' valleys[27,28,40]. This causes an out-of-plane tilt of the spin texture and should, in principle, reduce the in-plane component induced by the Rashba term[40], which arises from the perpendicular electric field at the interface due to broken inversion symmetry. Additionally, theoretical calculations based on realistic values show that the spin Hall angle of graphene/MoS$_2$ is at least one order of magnitude larger than the corresponding REE efficiency[31].

Even though the calculated spin Hall angle of 2.0% at 100 K and 1.7% at 300 K is smaller than in transition metals as Pt[61] or Ta[62] that have been used for graphene-based spintronic devices[63] or for spin-orbit torque magnetization switching[52], the maximum output signal $\Delta R_{SCC}$ of 209 mΩ is an order of magnitude larger than the maximum non-local ISHE signal reported for a graphene/metal device (11 mΩ at 300 K)[63] or for our recent graphene/TMD device (25 mΩ at 10 K)[38]. One major difference between the SCC in spin-orbit proximitized graphene and other devices is that transport of the spin current and conversion into a charge current happen in the same material, in the graphene channel itself, and no losses due to spin absorption across an interface or shunting occur.

However, $\Delta R_{SCC}$ is not a good figure of merit if one needs to compare efficiencies in the achievable output voltage across different materials and geometries in non-local devices. We recently proposed[38] an adjusted quantification of the conversion efficiency by defining the ratio $R_{eff}$, which has the units of resistance and is calculated by dividing the output voltage by the input spin current at the conversion region, that actually plays a role in the conversion. The advantage is that additional factors such as the spin polarization of the injector and the properties of the spin diffusion channel do not influence $R_{eff}$. In our case, it can be calculated using the following equation (see Note S9 for the calculation of the correction factor due to diffusive broadening in the precession):

$$R_{eff} = \frac{2\theta_{SH}^{eff} R_{sq}^{gr/TMD} \lambda_s^{eff}}{W_{cr}} \left(1 - e^{-W_{cr}/\lambda_s^{eff}}\right) \quad (1)$$

where $R_{sq}^{gr/TMD}$ is the square resistance of the proximitized region and $W_{cr}$ the width of the graphene Hall bar arm. The values for $R_{eff}$ at different representative temperatures and gate voltages are shown in Table 1 (all temperatures and gate voltages in Note S12 and S13). This normalized efficiency in our graphene/WSe$_2$ heterostructure (160 Ω at 100 K and -5 V) is eleven times larger than in our previously reported graphene/MoS$_2$ heterostructures (13.4 Ω)[38] and three orders of magnitude larger than in graphene/Pt-based devices (0.27 Ω)[63], using always the best case scenario.

|  | 300 K | 100 K (0 V) | 10 K | 100 K (-5 V) |
|---|---|---|---|---|
| $\Delta R_{SCC}$ (mΩ) | 38 ± 2 | 55 ± 1 | 90 ± 3 | 209 ± 1 |
| $\theta_{SH}^{eff}$ (%) | 1.7 ± 0.2 | 2.0 ± 0.2 | 2.8 ± 0.3 | 8.4 ± 0.2 |
| $\lambda_s^{eff}$ (nm) | 295 ± 14 | 380 ± 50 | 410 ± 50 | 480 ± 40 |
| $\theta_{SH}^{eff} \lambda_s^{eff}$ (nm) | 4.9 ± 0.7 | 7.6 ± 0.7 | 12 ± 2 | 41 ± 3 |
| $R_{eff}$ (Ω) | 21 ± 3 | 41 ± 4 | 65 ± 8 | 160 ± 12 |

**Table 1.** Spin-to-charge conversion parameters for selected temperatures and gate voltages. $\Delta R_{SCC}$ is the SCC signal and $\theta_{SH}^{eff}$ the spin Hall angle. However, the output current efficiency of a material is better quantified with the SCC length (the product of $\theta_{SH}^{eff}$ and $\lambda_s^{eff}$). To compare the output voltage efficiency across different devices, we calculate the normalized efficiency $R_{eff}$ with equation 1. The extracted parameters at other temperatures and gate voltages are listed in Notes S12 and S13, respectively.

If we are interested in the output current efficiency (for instance in the case of spin-orbit torques for magnetic switching), the $\theta_{SH}^{eff} \lambda_s^{eff}$ product (SCC length) is the proper figure of merit[9], which has units of length and compares straightforwardly with the Edelstein length that quantifies the efficiency of the IREE[59]. We obtained a SCC length up to an order of magnitude larger at room temperature (4.9 nm, or ~2.45 nm if we correct for a maximum overestimation of a factor of 2, as discussed in Note S11) than in the best heavy metals such as Pt[61] or Ta[62] (0.1-0.3 nm) or metallic interfaces such as Bi/Ag[64] (0.2-0.3 nm). MoTe$_2$, a semimetallic TMD, shows similar high efficiencies at room temperature (>1.15 nm)[65], slightly lower than the best results of topological insulators at room temperature (2.1 nm)[66]. Impressively, our maximum value of 41 nm at 100 K and -5 V back-gate voltage (see Table 1) is six times larger than the largest value reported so far, in the LAO/STO system (6.4 nm)[51] and still three times larger (~20 nm) if we assume the overestimation of our model (see Note S11).

Finally, it is also worth noting that, even though the SCC signal at 300 K is smaller than at low temperatures, the modulation due to the gate voltage could amplify it immensely as it is stronger than the temperature dependence of the signal (see Note S7). Applying higher negative gate voltages could also lead to giant ISHE signals at room temperature as we see from the charge transport measurements that the saturation region far away from the Dirac point is not reached yet (see Note S2).

We report for the first time SHE due to spin-orbit proximity in a graphene/WSe$_2$ van der Waals heterostructure. The temperature dependence of the spin transport and spin-to-charge conversion parameters are quantified, showing a robust performance up to room temperature. Interestingly, ISHE appears as the only SCC mechanism without an accompanying IREE, suggesting the dominance of the valley-Zeeman term over the Rashba term in the proximity-induced SOC. Additionally, we are able to directly gate control the SCC signal, tuning it from an off state up to 209 mΩ, while increasing the conversion efficiency. This leads to a very large $\theta_{SH}^{eff} \lambda_s^{eff}$ product above 20 nm in the best scenario (at 100 K and -5 V), with a remarkable 2.5 nm at room temperature and zero gate voltage. Our results demonstrate graphene/TMD as superior SCC material systems.

Note: After the completion of the current research, we became aware of recent results that show the electrical control of the SHE and the REE (with a SCC length of 3.75 nm and 0.42 nm respectively at room temperature) using WS$_2$[67] and the REE using the semimetal MoTe$_2$[68] and the metallic TaS$_2$[69] in proximity to graphene in van-der-Waals heterostructures.

## Supplementary Material

See the supplementary material for device fabrication and characterization, discussion of the missing low field crossing, additional antisymmetric and symmetric Hanle measurements, detailed information on the fitting to the diffusion equation, determination of the parameter uncertainties due to the effective model and tables with all extracted parameters.

## Acknowledgments

The authors thank Roger Llopis for drawing the sketches used in the figures. This work is supported by the Spanish MICINN under the Maria de Maeztu Units of Excellence Programme (MDM-2016-0618) and under projects MAT2015-65159-R, RTI2018-094861-B-100, and MAT2017-82071-ERC and by the European Union H2020 under the Marie Curie Actions (794982-2DSTOP and 766025-QuESTech). J.I.-A. acknowledges postdoctoral fellowship support "Juan de la Cierva - Formación" program by the Spanish MICINN (Grant No. FJC2018-038688-I). N.O. thanks the Spanish MICINN for a Ph.D. fellowship (Grant No. BES-2017-07963).

# Supplementary Material

# Gate tunability of highly efficient spin-to-charge conversion by spin Hall effect in graphene proximitized with WSe$_2$

Franz Herling[1], C. K. Safeer[1], Josep Ingla-Aynés[1], Nerea Ontoso[1], Luis E. Hueso[1,2], Fèlix Casanova[1,2,*]

[1]CIC nanoGUNE, 20018 Donostia-San Sebastian, Basque Country, Spain.
[2]IKERBASQUE, Basque Foundation for Science, 48013 Bilbao, Basque Country, Spain.

*E-mail: f.casanova@nanogune.eu

## Table of contents:

## S1. Device fabrication and electrical measurements

In a first step, graphene was mechanically exfoliated on a n-doped Si substrate with 300 nm of $SiO_2$ on top and a graphene flake of sufficient size was chosen. The number of layers of the graphene flake (three) was determined after the measurement with Raman spectroscopy (see Note S2). Secondly, a $WSe_2$ few-layer flake was placed on top of the trilayer graphene (TLG) flake by dry transfer with polydimethylsiloxane (PDMS). To pattern the TLG flake into a Hall bar structure, we used electron-beam lithography and reactive ion etching. Subsequently, we annealed the sample at 400 ºC at ultra-high vacuum to clean the surface of the flake and ensure a good interface between TLG and $WSe_2$. Electrical contacts were made by electron-beam lithography followed by electron-beam evaporation (5 nm of Ti) and thermal evaporation (40 nm of Au) in in a base pressure of $10^{-7}$ mbar. Next, four ferromagnetic (FM) electrodes were patterned by electron-beam lithography with varying widths to achieve different coercive fields. Afterwards, $TiO_x$ tunnel barriers were created (2.4 Å of e-beam evaporated Ti and oxidation in air), and 35 nm of Co were e-beam evaporated. The resulting contact resistance of the $Co/TiO_x$ electrodes is between 5 and 50 kΩ. The final device can be seen in Figure 1b of the main text and in more detail in Note S2.

The sample is wire bonded to a chip carrier and placed in a physical property measurement system by Quantum Design. All electrical measurements are performed between 10 K and 300 K using a direct-current reversal technique to exclude heating effects employing a Keithley 2182A nanovoltmeter and a 6221 current source (10 µA). The n-doped Si substrate acts as a back-gate electrode to which we apply the gate voltage with a Keithley 2636B (compliance set to 1 nA). The sample holder can be rotated along two planes in the magnetic field of the superconducting solenoid magnet allowing us to apply a magnetic field in $x$- and $y$-direction. An initial standard electrical characterization of the device can be found in Note S2.

## S2. Characterization of the final device after the electrical measurements

After finishing the electrical measurements, the final device was imaged by scanning electron microscopy to determine the lateral dimensions (see Figure S1), atomic force microscopy to determine the thickness of the $WSe_2$ flake (see Figure S2) and Raman microscopy to determine the thickness of the graphene flake (see Figure S3). This was only done after the fabrication and electrical measurements to minimize the exposure to atmosphere and limit the contamination and degradation of the sample.

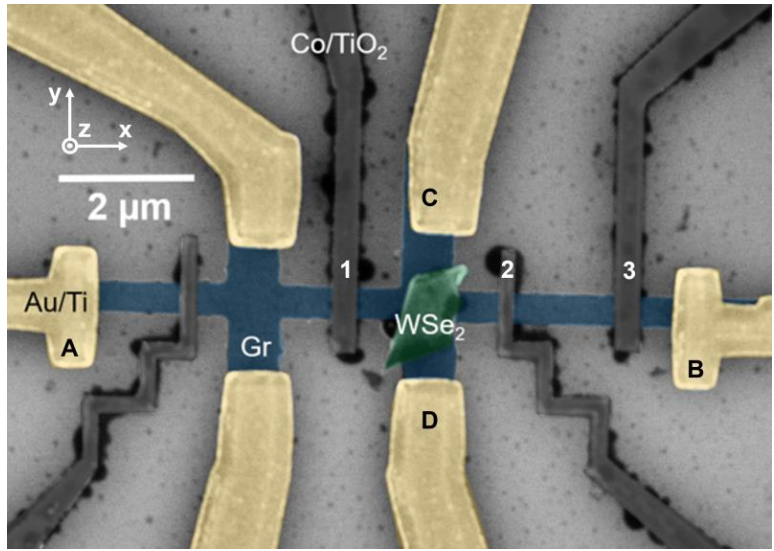

**Figure S1.** False-colored scanning electron microscopy image of the device after the electrical measurements. The oxidation of the Co electrodes is visible as well as some contamination. The width of the graphene channel was measured as 495 nm, the width of the graphene Hall bar arms is 810 nm. The center-to-center distance between the Co electrodes is 1.84 µm for the reference lateral spin valve (LSV) on the right (electrode 2 and 3) and 2.48 µm for the graphene/$WSe_2$ LSV in the middle (electrode 1 and 2). The distance from the left edge of the Hall bar arms on the right (graphene/$WSe_2$ region) to the center of the FM electrode 1 is 870 nm.

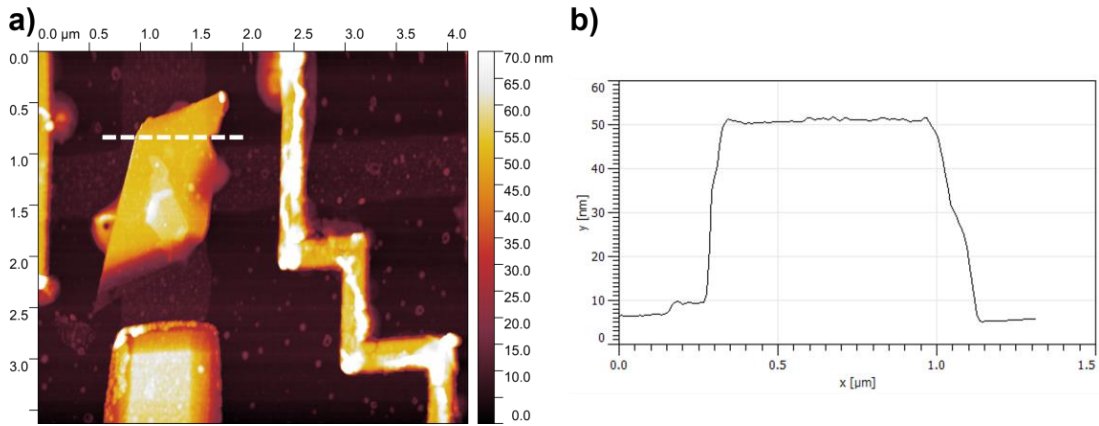

**Figure S2.** Atomic force microscopy characterization of the device after the electrical measurements. **a)** Area scan in tapping mode showing the topography of the device. The oxidation of the Co electrodes is visible as well as some contamination. **b)** Line profile taken along the marked line across the $WSe_2$ flake. The thickness of the $WSe_2$ flake is 45 nm.

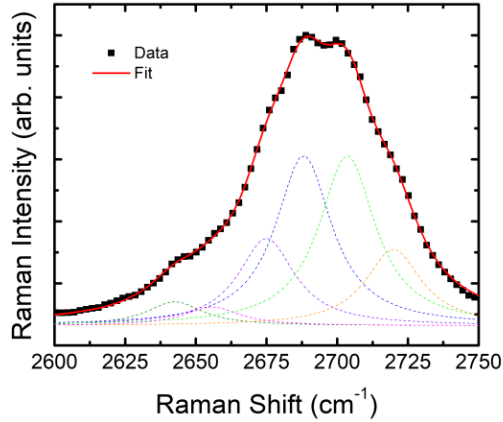

**Figure S3.** Raman spectroscopy of the graphene channel after the electrical measurements with a confocal microscope using a green polarized laser (532 nm). The Raman intensity profile can be fitted with 6 Lorentzian functions with Full Width Half Maximum of 24 cm$^{-1}$, which gives a thickness of three layers for the graphene flake[1].

## S3. Electrical characterization of the device

To electrically characterize the device before the non-local measurements, we measured the four-point resistance of the graphene channel between the electrodes of the reference graphene LSV and of the graphene/WSe$_2$ LSV and calculated the corresponding sheet resistance $R_{sq}$. The temperature dependence of $R_{sq}$ for the two regions is shown in Figure S4a. The transition metal dichalcogenide (TMD) not only enhances the spin-orbit coupling (SOC) by proximity effect, but also dopes the graphene. Figure S4b shows the gate dependence of both regions at 100 K and how the different doping moves the Dirac point from higher positive gate voltages to between 5 and 6 V. Due to a larger leakage current (> 1 nA) we were not able to apply higher gate voltages so that a full analysis of the charge transport measurements was not possible. One reason for this could be damage to the SiO$_2$ from the wire bonding.

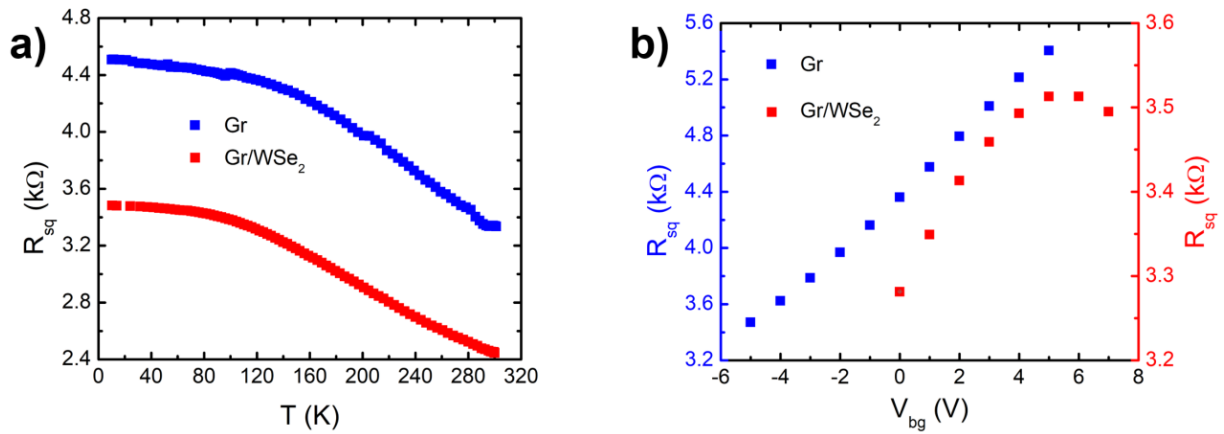

**Figure S4.** Sheet resistance characterization of the reference graphene LSV and the graphene/WSe$_2$ LSV. Measured by applying a current along the main graphene channel (along the $x$-axis) with the Au/Ti contacts A and B and using the FM electrodes pairs (2 and 3 for the pristine graphene LSV, 1 and 2 for the graphene/WSe$_2$ LSV) as a voltage probe. Sheet resistance for both LSVs as a function of **a)** temperature at 0 V back-gate voltage and **b)** applied back-gate voltage at 100 K.

## S4. Discussion of the missing low field crossing in the Hanle measurements for the graphene/WSe₂ LSV

Some factors can influence the measurement of the non-local resistance shown in Figure 2d. Firstly, we note that this cannot be a background-related effect as both curves overlap within the noise level when the magnetizations have saturated. Secondly, an out-of-alignment mounting of the sample in the rotational sample holder should not affect our measurements as it does not change the orientation of the electrodes to each other in regard to the magnetic field. Furthermore, such a misalignment must be very small in our measurements, as otherwise it would lead to premature switching of the contact magnetizations when $B_x$ is applied but this is not visible in our data. A slight out-of-plane misalignment between the two FM electrodes due to inhomogeneous magnetic domain formation could lead to a shift of the Hanle curves. However, this effect would shift the Hanle precession data with respect to $B_x$ and would not impede the crossing of both curves at low field. We also note that the interpretation of the data could be more complex due to local invariances of the strength of the SOC, as the proximity effect can depend on the distance between the two flakes[2] and could vary due to wrinkles or strain after the stamping, that would have to persist after annealing. However, this cannot affect the sign of the spin signal itself unless the Landé factors would change sign, leading to complex precession processes. Hence, we cannot determine the reason for the missing crossing of $R_{NL}^P$ and $R_{NL}^{AP}$ in Figure 2d.

## S5. Antisymmetric Hanle curve measurement for direct spin Hall effect across the graphene/WSe₂ region

Swapping the contact pairs of the inverse spin Hall effect (ISHE) measurement, shown in the main text in Figure 3, enables us to directly observe the spin Hall effect (SHE). In this case, we apply a charge current across the graphene/WSe$_2$ region. Due to the proximity-induced SHE, a spin current diffuses along the graphene channel (along the $x$-axis) with out-of-plane spins. An in-plane magnetic field applied along the $x$-axis ($B_x$) precesses the spins towards the $y$-axis, which can then be detected with the FM electrode. The measurement is shown in Figure S5a. The charge-to-spin conversion signal is only slightly smaller than its Onsager reciprocal, but the measurement is noisier as it uses the FM electrode for detection that has a higher contact resistance due to the TiO$_x$ tunnel barrier than the Au/Ti contacts. Therefore, all measurements in the main text were performed in the ISHE setup.

Figure S5b compares $R_{SCC}$ for the SHE and ISHE measurement. The opposite precession in the antisymmetric Hanle curve is expected for the direct SHE measurement and is therefore another strong evidence for SHE due to proximity effect in our device.

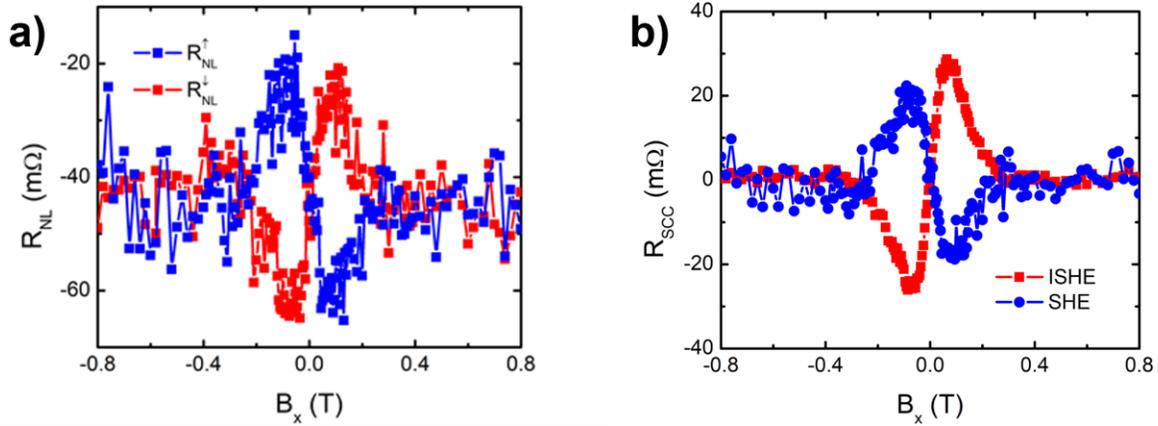

**Figure S5.** Charge-to-spin conversion measurement at 100 K. **a)** Non-local charge-to-spin conversion curves obtained by applying a charge current across the graphene/WSe$_2$ Hall bar and measuring between FM electrode and Au contact (current C-D and voltage 1-A in Figure S1). A magnetic field is applied along the in-plane hard axis direction ($B_x$) for initial positive ($R_{NL}^{\uparrow}$, blue) and negative ($R_{NL}^{\downarrow}$, red) magnetization directions of the FM electrode. **b)** Comparison of $R_{SCC}$ for the SHE (current C-D and voltage 1-A) and ISHE (current 1-A and voltage C-D) measurements as a function of an in-plane magnetic field ($B_x$).

## S6. Antisymmetric Hanle curve measurement for inverse spin Hall effect from the right side across the graphene/WSe$_2$ region

Additionally, we measured the ISHE by injecting the spin current from both sides of the graphene/WSe$_2$ region. The antisymmetric Hanle curve for injection from the right side can be seen in Figure S6a. The measurement has a similar signal amplitude to the one from the left side (Figure 3a of the main text) but larger noise and a linear background due to drift. Therefore, all measurements in the main text were performed with electrode 1.

Figure S6b compares $R_{SCC}$ for the ISHE measurements injecting spin current from the left and the right side of the graphene/WSe$_2$ region. Again, the opposite precession in the antisymmetric Hanle curve is expected for injecting a spin current from the opposite direction into the graphene/WSe$_2$ region and is therefore another strong evidence for SHE due to proximity effect in our device.

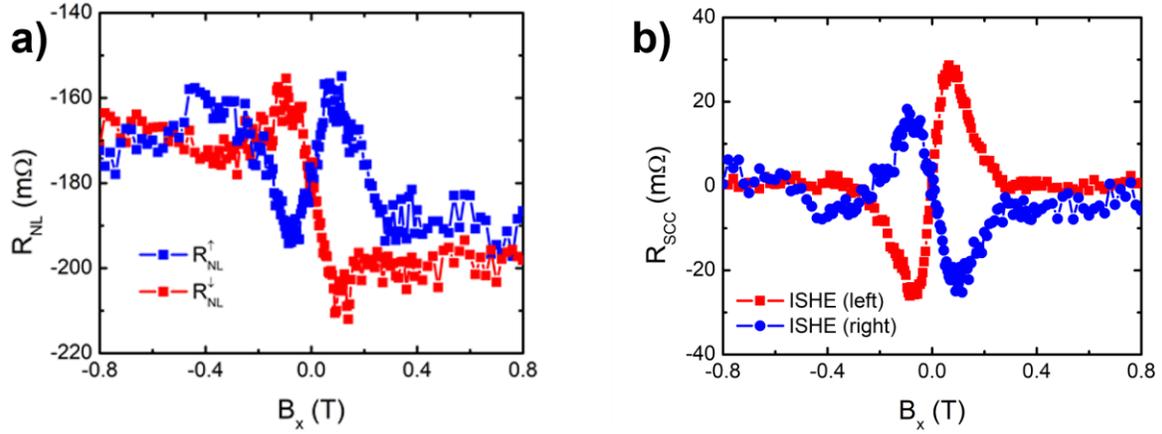

**Figure S6.** Spin-to-charge conversion (SCC) measurement at 100 K. **a)** Non-local SCC conversion curves obtained by measuring across the graphene/WSe$_2$ Hall bar (current 2-B and voltage C-D in Figure S1), therefore injecting a spin current from the right side of the proximitized region. A magnetic field is applied along the in-plane hard axis direction ($B_x$) for initial positive ($R_{NL}^\uparrow$, blue) and negative ($R_{NL}^\downarrow$, red) magnetization directions of the FM electrodes. **b)** Comparison of $R_{SCC}$ for the ISHE measurements from the left (current 1-A and voltage C-D) and right side (current 2-B and voltage C-D) of the graphene/WSe$_2$ region as a function of an in-plane magnetic field ($B_x$).

## S7. Symmetric Hanle curve measurements across the reference graphene lateral spin valve at different temperatures and gate voltages

Complementary to the measurements shown in Figure 4 of the main text, we also measured the symmetric Hanle curves for all the shown temperatures and gate voltages. Representative curves are shown alongside their fits to the diffusion equation in Figure S7. The non-local spin signal in the reference graphene LSV between electrode 2 and 3 decreases for lower temperatures as seen in the smaller peak amplitudes in Figure S7a. For the gate voltage modulation, shown in Figure S7b, no clear trend is observable.

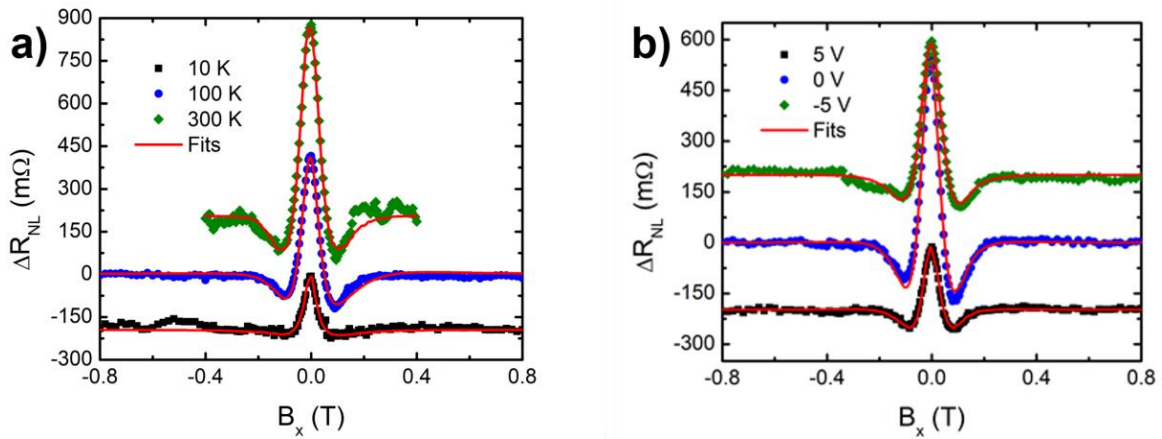

**Figure S7.** Net symmetric Hanle signals measured at **a)** different temperatures and zero back-gate voltage and **b)** different back-gate voltages and 100 K. Additional measurements and fits at 50 K, 150 K, 200 K and 250 K and at -3 V and 2 V are not shown here. The scatter plots are the experimental data, the red solid lines are fits to the data. Curves are shifted in the vertical axis for clarity.

## S8. Comparison of temperature and gate dependence of the spin-to-charge conversion signal

In Figure S8, we plot the SCC signal $\Delta R_{SCC}$ as a function of temperature for all measured temperatures in the range between 10 and 300 K and the $\Delta R_{SCC}$ values we measured at 100 K for all the back-gate voltages in the range from -5 V to 5 V. The increase with decreasing temperature is clearly visible but not as pronounced as the increase for negative gate voltages.

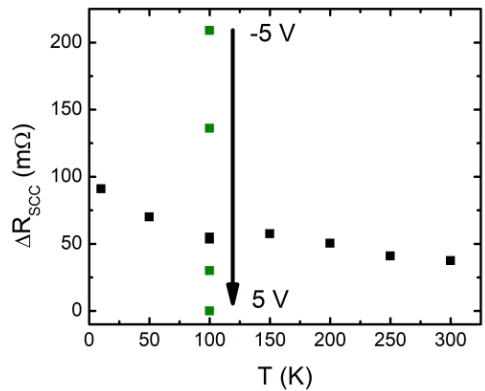

**Figure S8.** The SCC signal $\Delta R_{SCC}$ across the graphene/WSe$_2$ region as a function of temperature and the interval of the signal modulation by the gate voltage at 100 K. Gate voltage measurements (green) for -5 V, -3 V, 2 V and 5 V in the order of the arrow.

## S9. Pulling of the magnetization of the ferromagnetic electrode

When fitting the diffusion equation to the experimental data of the Hanle precession, one has to account for the signal injected parallel to $B_x$ due to the pulling of the magnetization of the injecting FM electrode in field direction[3]. Therefore, we calculated the angle $\beta$ between magnetization of the electrode and the field for all the fits to the symmetric Hanle curves for different temperatures and back-gate voltages. One curve is shown exemplary in Figure S9. The calculation of $\beta$ and how to include it in the fitting is explained in detail in Ref. 2 (Note S1.1 of the Supporting Information).

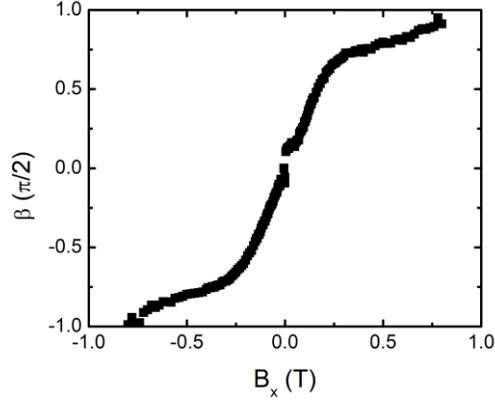

**Figure S9.** Angle $\beta$ between the FM electrode magnetization and the easy axis extracted from the symmetric Hanle data in Figure 2b of the main text as a function of an applied in-plane magnetic field at 100 K for no applied back-gate voltage.

### S10. Correction factor for $R_{eff}$ due to diffusive broadening

The normalized efficiency $R_{eff}$ represents the SCC output signal for injecting a spin current with an out-of-plane polarization without the influence of additional factors such as the spin polarization of the injector and the properties of the spin diffusion channel (see equation 1 in the main text). However, in our device this is achieved by injecting spins polarized along the $y$-axis and precessing them along the $z$-axis by applying a magnetic field along the spin current direction ($B_x$). The diffusive transport of the spins introduces a broadening of the precession angle and therefore lowers the signal. To compensate for this and compare our results with experiments without precession[4], we calculated both SCC signals with the spin transport and SCC parameters extracted at 100 K and 0 V back-gate voltage. We estimated the correction factor here with 76 % using the values in Figure S10. The correction factor is also discussed in Ref. 2 (Note S7 of the Supporting Information).

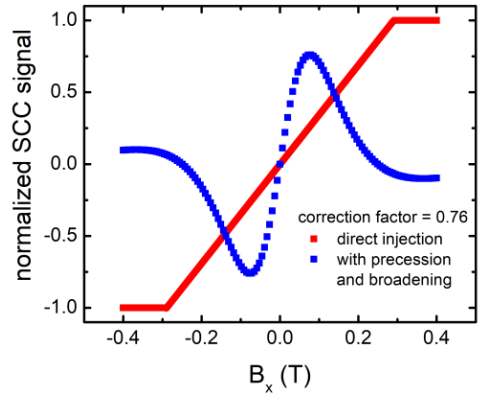

**Figure S10.** Calculation of the correction factor for the normalized conversion efficiency $R_{eff}$ due to diffusive broadening of the spin precession. Equation S2 in Ref. 2 was used to calculate the SCC signal for a precessing (blue) and a directly out-of-plane injected (red) spin current as a function of an in-plane

magnetic field ($B_x$) with the parameters extracted from the fitting of the symmetric and antisymmetric Hanle curves. The ratio between the maximum of the blue curve and the constant red curve gives the correction factor of around 76 %.

## S11. Comparison of the gate dependence of the spin-to-charge conversion signal and the sheet resistance

To show that the increase of the measured SCC signal $\Delta R_{SCC}$ with back-gate voltage is not simply due to an increase in the sheet resistance $R_{sq}$ of the graphene/WSe₂ region, but due to a more effective SCC, we plotted $\Delta R_{SCC}$ and $R_{sq}^{gr/TMD}$ as a function of back-gate voltage $V_{bg}$ in Figure S11. $\Delta R_{SCC}$ at 0 V is amplified by 400 % with applying -5 V and turned off with applying 5 V back-gate voltage while $R_{sq}^{gr/TMD}$ of the graphene/WSe₂ region changes in the opposite direction.

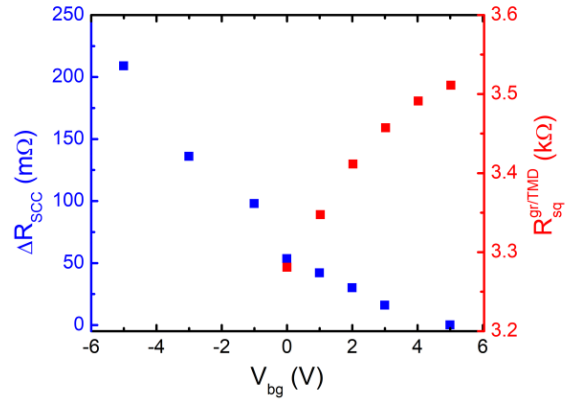

**Figure S11.** The SCC signal $\Delta R_{SCC}$ and the sheet resistance $R_{sq}^{gr/TMD}$ of the graphene/WSe₂ region as a function of an applied back-gate voltage $V_{bg}$ at 100 K.

## S12. Determination of the parameter uncertainties due to the homogeneous model simplification

The spin transport parameters relevant to determine $\theta_{SH}^{eff}$ in our device have been obtained using the one-region model that assumes that the spin transport parameters of the pristine graphene region between electrode and Hall bar arms are the same as those of the graphene/WSe₂ region (see Note S1.1 of the Supporting Information of Ref. 2). However, we know that this assumption leads to uncertainties. In particular, it leads to the extraction of a spin lifetime ($\tau_s^{eff}$), which is an average of ($\tau_s^{gr}$) and the spin lifetime of the graphene/TMD region ($\tau_s^{gr/TMD}$). Although this spin lifetime is anisotropic $and\ \tau_\perp^{gr/TMD} > \tau_\parallel^{gr/TMD}$, $\tau_\perp^{gr/TMD}$ is typically shorter than $\tau_s^{gr}$[5,6]. Hence, the use of $\tau_s^{eff}$ leads to the overestimation of $\tau_\perp^{gr/TMD}$ and the extracted $\theta_{SH}^{eff}$ is underestimated by our model.

However, due to the complexity of the system, the actual magnitude of the underestimation cannot be obtained from simple considerations. To disentangle it, we have modelled a geometry that accounts for the different spin transport properties of the TMD-covered and the pristine graphene regions by dividing our channel in 4 different regions (see inset of Figure S12a). Region 1 is at the left side of the spin injector and is semi-infinite, region 2 connects the spin injector and the TMD-proximitized graphene region, which is region 3. Finally, we add a pristine graphene region (region 4) which is placed at the right side of region 3.

To determine the spin accumulations in our device, we model the spin propagation using the Bloch equations:

$$D_s^{gr(gr/TMD)} \frac{d^2\vec{\mu}}{dx^2} - \frac{\vec{\mu}}{\tau_{s(\perp)}^{gr\left(\frac{gr}{TMD}\right)}} - \vec{\omega} \times \vec{\mu} = 0$$

Here, $\vec{\mu} = (\mu_{s_x}, \mu_{s_y}, \mu_{s_z})$ is the spin accumulation, $\vec{\omega} = g\mu_B \vec{B}$ is the Larmor frequency, $g = 2$ the Landé factor, $\mu_B$ the Bohr magneton, and $\vec{B} = (B_x, B_y, B_z)$ the applied magnetic field. When a magnetic field is applied along the $x$-direction, it induces spin precession in the $y - z$ plane and the solution to the Bloch equation is:

$$\mu_{s_y} = A\exp\left(\frac{x}{\lambda_{s(\perp)}^{gr(gr/TMD)}}\sqrt{1 + i\omega\tau_{s(\perp)}^{gr(gr/TMD)}}\right) + B\exp\left(\frac{x}{\lambda_{s(\perp)}^{gr(gr/TMD)}}\sqrt{1 - i\omega\tau_{s(\perp)}^{gr(gr/TMD)}}\right)$$

$$+ C\exp\left(-\frac{x}{\lambda_{s(\perp)}^{gr(gr/TMD)}}\sqrt{1 + i\omega\tau_{s(\perp)}^{gr(gr/TMD)}}\right)$$

$$+ D\exp\left(-\frac{x}{\lambda_{s(\perp)}^{gr(gr/TMD)}}\sqrt{1 - i\omega\tau_{s(\perp)}^{gr(gr/TMD)}}\right)$$

$$\mu_{s_z} = -iA\exp\left(\frac{x}{\lambda_+^{gr(gr/TMD)}}\right) + iB\exp\left(\frac{x}{\lambda_-^{gr(gr/TMD)}}\right) - iC\exp\left(-\frac{x}{\lambda_+^{gr(gr/TMD)}}\right)$$

$$+ iD\exp\left(-\frac{x}{\lambda_-^{gr(gr/TMD)}}\right)$$

where $\tau_{s(\perp)}^{gr(gr/TMD)}$, $D_s^{gr(gr/TMD)}$ and $R_{sq}^{gr(gr/TMD)}$ are the spin lifetime, diffusion coefficient and square resistance of the pristine (graphene/WSe$_2$) region and the associated spin relaxation lengths are obtained using $\lambda_s = \sqrt{D_s \tau_s}$. $\lambda_{s(\perp)}^{gr(gr/TMD)}/\lambda_{\pm}^{gr(gr/TMD)} = \sqrt{1 \pm i\omega\tau_{s(\perp)}^{gr(gr/TMD)}}$ and $A$, $B$, $C$, and $D$ are coefficients determined by the boundary conditions that depend on the device geometry. Note that, because $\tau_{\perp}^{gr/TMD} \neq \tau_{\parallel}^{gr/TMD}$, the spin precession in the TMD-covered region is better described by the solution to the anisotropic Bloch equations. However, in our case, because $\tau_{\perp}^{gr/TMD} < \tau_s^{gr}$ and the inverse spin Hall detection at the TMD-proximitized graphene region starts at the left edge of this region (see Figure S12a), precession in this region becomes less relevant and can be simplified taking $\lambda_\perp^{gr/TMD} = \sqrt{D_s \tau_\perp^{gr/TMD}}$. $\tau_\perp^{gr/TMD}$ is the relevant spin lifetime here because the converted spins point along $z$ due to the symmetry of the ISHE. We have confirmed this by simulating the anisotropic system and obtained a small deviation of about 10 % in the extracted parameters for a typical anisotropy ($\tau_\perp^{gr/TMD}/\tau_\parallel^{gr/TMD} = 10$).

The model sketched in the inset of Figure S12a has 4 different regions. Region 1 extends from $x \to -\infty$ to the spin injector, region 2 connects the spin injector with the TMD-covered region, which is region 3. Finally, region 4 is placed at the right of region 3 and extends until $x \to \infty$. Regions 1, 2 and 4 are pristine graphene and 3 is proximitized by WSe$_2$.

To determine the relevant parameters ($A$, $B$, $C$, and $D$) in the different regions of our device we use the following boundary conditions:
1. The spin accumulation $\mu_s \to 0$ when $x \to \pm\infty$.
2. The spin accumulation is continuous everywhere.
3. The spin currents are defined as $I_{S_{y(z)}}^{gr(gr/TMD)} = -\frac{W_{gr}}{eR_{sq}^{gr(gr/TMD)}} \frac{d\mu_{s_{y(z)}}}{dx}$ for spins pointing in the $y(z)$-direction in the graphene and are continuous at the intersection between all the regions apart from the spin injection point. $W_{gr}$ is the width of the graphene channel.
4. $I_s^{gr}$ has a discontinuity at the spin injector ($x = 0$) of $\Delta I_{S_y}^{gr} = P_i I_C - \frac{\mu_{s_y}(x=0)}{eR_{c1}}$ for spins pointing along $y$ and $\Delta I_{S_z}^{gr} = -\frac{\mu_{s_z}(x=0)}{eR_{c1}}$ for spins along $z$. Here $P_i$ and $R_{c1}$ are the spin polarization and contact resistance of the spin injector and $I_C$ is the applied charge current.

The SCC signal is obtained using:

$$R_{SCC} = \theta_{SH}^{gr/TMD} R_{sq} x_{shunt} \bar{I}_{s_z} / I_c \tag{S1}$$

Here, $\theta_{SH}^{gr/TMD}$ is the spin Hall angle associated with the conversion of spins pointing along $z$ and propagating along $x$. $x_{shunt} = 1$ is the shunting factor associated with the role of the WSe$_2$ as a parallel channel that reduces the effective resistance of the graphene, and $\bar{I}_{s_z}$ is the average spin current in the TMD-covered graphene region and is calculated using: $\bar{I}_{s_z} = \frac{1}{W_{cr}} \int_L^{L+W_{cr}} I_{s_z}(x) dx$. Additionally, $L$ is the distance between the center of the spin injector and the left side of the TMD-covered region and $W_{cr}$ is the width of the latter region.

The 1-region model used to fit the data assumes that the spin transport properties of region 3 are the same as 1, 2 and 4 and has the following solution[3]:

$$R_{SCC} = \frac{\theta_{SH}^{eff} R_{sq}^{gr/TMD} \bar{I}_{s_z}}{I_C} = \pm \frac{\theta_{SH}^{eff} P_i R_{sq}^{gr/TMD} \lambda_s^{eff}}{2W_{cr}} Im \left\{ \frac{e^{-\frac{L}{\lambda_s^{eff}}\sqrt{1-i\omega\tau_s^{eff}}}}{\sqrt{1-i\omega\tau_s^{eff}}} - \frac{e^{-\frac{L+W_{cr}}{\lambda_s^{eff}}\sqrt{1-i\omega\tau_s^{eff}}}}{\sqrt{1-i\omega\tau_s^{eff}}} \right\} \tag{S2}$$

Where the $\pm$ stands for the up and down magnetization configurations of the spin injector. To determine the effect of the 1-region approximation on the extracted parameters, we have used the relevant geometrical ($L = 0.87$ µm, $W_{cr} = 0.81$ µm and $W_{gr} = 0.50$ µm, see Figure S1) and transport parameters of the pristine graphene region (see Table S2 for $V_{bg} = 0$ V and T=100 K) as inputs to generate the antisymmetric Hanle precession curves from equation S1 assuming that $D_s^{gr/TMD} = D_s^{gr}$ and sweeping $B_x$ fields up to $\pm 1$ T. Each curve has been fitted to equation S2 (see Figure S12a). This operation has been performed for $\tau_\perp^{gr/TMD}$ ranging from 1 ps to 1 ns. Note that the models used here assume no contact pulling to obtain the most accurate estimate of $\tau_s^{eff}$, $D_s^{eff}$ and $\theta_{SH}^{eff}$. $\lambda_s^{eff}$ and $\theta_{SH}^{eff}$ extracted from the fits are plotted in Figure S12b normalized by the input $\lambda_\perp^{gr/TMD}$ and $\theta_{SH}^{gr/TMD}$. From this plot we observed that, in the low $\tau_\perp^{gr/TMD}$ range, despite the factor 9 overestimation in $\lambda_s^{eff}$, the $\theta_{SH}^{eff} \lambda_s^{eff}$ product is only overestimated by a factor 2.

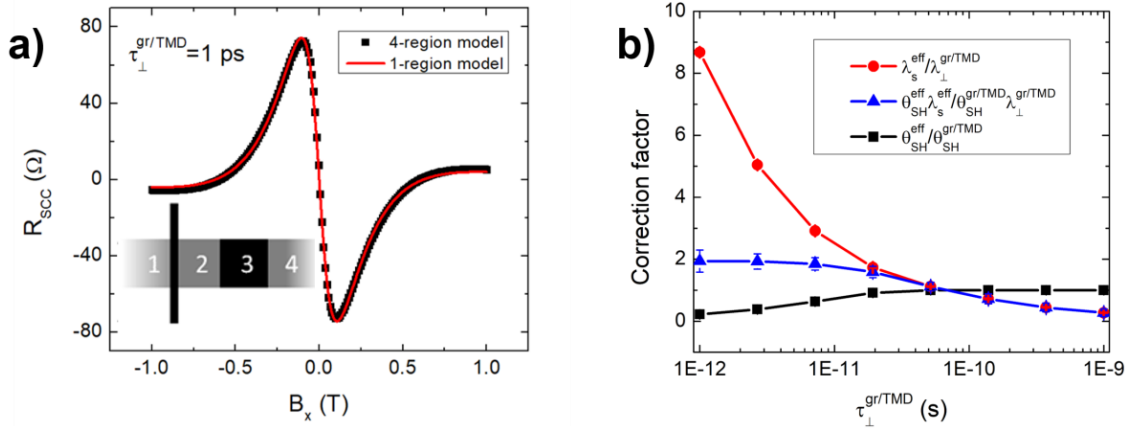

**Figure S12.** Parameter uncertainties induced by the 1-region model approximation. **a)** Antisymmetric Hanle curve ($R_{SCC}$) simulated from equation S1 using the 4-region model, with $\tau_\perp^{gr/TMD} = 1$ ps, $\tau_s^{gr} = 60$ ps and an arbitrary injection polarization $P_i = 1$ (black squares). This simulation is fitted to equation S2 of the 1-region model (red line). The inset at the bottom left corner is the 4-region model schematic, where the grey regions (1,2,4) are pristine graphene, the TMD-covered region (3) is black, and the FM electrode used for injection is represented by the black vertical bar. **b)** Correction factor between $\lambda_\perp^{gr/TMD}$ and $\lambda_s^{eff}$ (red circles), $\theta_{SH}^{gr/TMD}$ and $\theta_{SH}^{eff}$ (black squares) and their products (blue triangles), extracted with the procedure shown in panel a for different values of $\tau_\perp^{gr/TMD}$. Note that, when $\tau_\perp^{gr/TMD} = \tau_s^{gr}$, the correction factors are 1. The lines are a guide to the eye.

Note that our 4-region model does not account for spin diffusion into the top and bottom arms of the TMD-covered region used for the Hall measurements. Hence, because the spins propagating along $y$ do not contribute to the SCC voltage along $y$, $\theta_{SH}^{eff}$ is underestimated by both models, which again would decrease the correction factor calculated in this section.

## S13. Extracted parameters for different temperatures

| | 10 K | 50 K | 100 K | 150 K | 200 K | 250 K | 300 K |
|---|---|---|---|---|---|---|---|
| $R_{sq}^{gr}$ (Ω) | 4508 | 4467 | 4361 | 4260 | 3973 | 3641 | 3338 |
| $R_{sq}^{gr/TMD}$ (Ω) | 3482 | 3461 | 3281 | 3181 | 2905 | 2638 | 2445 |
| $D_s^{gr}$ ($10^{-3}$ m²/s) | 4.5 ± 0.3 | 4.3 ± 0.1 | 4.1 ± 0.1 | 5.5 ± 0.1 | 8.4 ± 0.2 | 4.8 ± 0.1 | 6.0 ± 0.2 |
| $\tau_s^{gr}$ (ps) | 127 ± 6 | 125 ± 2 | 90 ± 1 | 100 ± 2 | 108 ± 3 | 84 ± 2 | 92 ± 2 |
| $\lambda_s^{gr}$ (nm) | 760 ± 90 | 730 ± 30 | 610 ± 20 | 740 ± 30 | 950 ± 50 | 635 ± 30 | 740 ± 40 |
| $P_i$ (%) | 2.6 ± 0.1 | 3.5 ± 0.1 | 5.4 ± 0.1 | 4.1 ± 0.1 | 3.4 ± 0.1 | 8.5 ± 0.3 | 5.8 ± 0.2 |
| $D_s^{eff}$ ($10^{-3}$ m²/s) | 2.1 ± 0.1 | 2.2 ± 0.1 | 2.3 ± 0.2 | 2.3 ± 0.1 | 2.4 ± 0.1 | 1.9 ± 0.1 | 2.0 ± 0.1 |
| $\tau_s^{eff}$ (ps) | 81 ± 5 | 64 ± 6 | 61 ± 3 | 54 ± 1 | 42 ± 2 | 44 ± 2 | 44 ± 2 |
| $\lambda_s^{eff}$ (nm) | 410 ± 50 | 380 ± 35 | 380 ± 50 | 350 ± 17 | 320 ± 28 | 290 ± 11 | 295 ± 14 |
| $P_i^{eff}$ (%) | 2.7 ± 0.1 | 2.7 ± 0.1 | 3.3 ± 0.2 | 2.9 ± 0.1 | 3.4 ± 0.1 | 3.7 ± 0.2 | 3.1 ± 0.3 |
| $\Delta R_{SCC}$ (mΩ) | 90 ± 3 | 70 ± 2 | 55 ± 1 | 58 ± 1 | 51 ± 2 | 41 ± 1 | 38 ± 2 |
| $\theta_{SH}^{eff}$ (%) | 2.8 ± 0.3 | 2.0 ± 0.2 | 2.0 ± 0.2 | 2.1 ± 0.2 | 3.4 ± 0.2 | 1.6 ± 0.2 | 1.7 ± 0.2 |
| $\theta_{SH}^{eff}\lambda_s^{eff}$ (nm) | 12 ± 2 | 7.5 ± 0.9 | 7.6 ± 0.7 | 7.2 ± 0.4 | 11 ± 0.9 | 4.6 ± 0.5 | 4.9 ± 0.7 |
| $R_{eff}$ (Ω) | 65 ± 8 | 24 ± 4 | 41 ± 4 | 39 ± 2 | 54 ± 4 | 22 ± 2 | 21 ± 3 |

**Table S1.** Charge and spin transport parameters for the temperature range from 10 K to room temperature. $D_s^{gr}$, $\tau_s^{gr}$ and $P_i$ are extracted from the fits to the symmetric Hanle precession curves, $D_s^{eff}$, $\tau_s^{eff}$ and $P_i^{eff}$ from the antisymmetric ones. They enable us to calculate $\lambda_s^{gr}$, $\lambda_s^{eff}$ and $\theta_{SH}^{eff}$. $\Delta R_{SCC}$ is the spin-to-charge conversion signal and $R_{eff}$ the normalized conversion efficiency.

## S14. Extracted parameters for different gate voltages

| | -5 V | -3 V | 0 V | 2 V | 5 V |
|---|---|---|---|---|---|
| $R_{sq}^{gr}$ (Ω) | 3470 | 3788 | 4361 | 4794 | 5405 |
| $R_{sq}^{gr/TMD}$ (Ω) | (2574) | (2810) | 3281 | 3413 | 3513 |
| $D_s^{gr}$ ($10^{-3}$ m²/s) | 7.1 ± 0.1 | 4.9 ± 0.1 | 4.1 ± 0.1 | 4.0 ± 0.1 | 3.6 ± 0.1 |
| $\tau_s^{gr}$ (ps) | 100 ± 2 | 88 ± 1 | 90 ± 1 | 103 ± 2 | 106 ± 2 |
| $\lambda_s^{gr}$ (nm) | 840 ± 30 | 660 ± 20 | 610 ± 20 | 640 ± 30 | 620 ± 30 |
| $P_i$ (%) | 2.9 ± 0.1 | 4.6 ± 0.2 | 5.4 ± 0.1 | 5.9 ± 0.2 | 3.4 ± 0.1 |
| $D_s^{eff}$ ($10^{-3}$ m²/s) | 3.5 ± 0.1 | 3.5 ± 0.2 | 2.3 ± 0.2 | 2.1 ± 0.1 | - |
| $\tau_s^{eff}$ (ps) | 67 ± 4 | 45 ± 5 | 61 ± 3 | 68 ± 2 | - |
| $\lambda_s^{eff}$ (nm) | 480 ± 40 | 400 ± 60 | 380 ± 50 | 280 ± 25 | - |
| $P_i^{eff}$ (%) | 4.9 ± 0.1 | 4.2 ± 0.2 | 3.3 ± 0.2 | 2.8 ± 0.1 | - |
| $\Delta R_{SCC}$ (mΩ) | 209 ± 1 | 136 ± 1 | 55 ± 1 | 30 ± 1 | 0 ± 2 |
| $\theta_{SH}^{eff}$ (%) | 8.4 ± 0.2 | 3.8 ± 0.2 | 2.0 ± 0.2 | 1.3 ± 0.1 | <0.2 % |
| $\theta_{SH}^{eff} \lambda_s^{eff}$ (nm) | 41 ± 3 | 15 ± 2 | 7.6 ± 0.7 | 3.7 ± 0.3 | - |
| $R_{eff}$ (Ω) | 160 ± 12 | 70 ± 9 | 41 ± 4 | 23 ± 2 | - |

**Table S2.** Charge and spin transport parameters for the temperature range from 10 K to room temperature. $D_s^{gr}$, $\tau_s^{gr}$ and $P_i$ are extracted from the fits to the symmetric Hanle precession curves, $D_s^{eff}$, $\tau_s^{eff}$ and $P_i^{eff}$ from the antisymmetric ones. They enable us to calculate $\lambda_s^{gr}$, $\lambda_s^{eff}$ and $\theta_{SH}^{eff}$. $\Delta R_{SCC}$ is the spin-to-charge conversion signal and $R_{eff}$ the normalized conversion efficiency. The values of $R_{sq}^{gr/TMD}$ for -3 V and -5 V are determined by a linear fit to the data left of the Dirac point shown in Figure S4b.